# BAMHealthCloud: A Biometric Authentication and Data Management System for Healthcare Data in Cloud

Kashish A. Shakil*, *Student Member, IEEE*, Farhana J. Zareen*, Mansaf Alam, *Member, IEEE*

And Suraiya Jabin

*Abstract-* Advancements in healthcare industry with new technology and population growth has given rise to security threat to our most personal data. The healthcare data management system consists of records in different formats such as text, numeric, pictures and videos leading to data which is big and unstructured. Also, hospitals have several branches at different locations throughout a country and overseas. In view of these requirements a cloud based healthcare management system can be an effective solution for efficient health care data management. One of the major concerns of a cloud based healthcare system is the security aspect. It includes theft to identity, tax fraudulence, insurance frauds, medical frauds and defamation of high profile patients. Hence, a secure data access and retrieval is needed in order to provide security of critical medical records in health care management system. Biometric authentication mechanism is suitable in this scenario since it overcomes the limitations of token theft and forgetting passwords in conventional token id-password mechanism used for providing security. It also has high accuracy rate for secure data access and retrieval. In this paper we propose BAMHealthCloud which is a cloud based system for management of healthcare data, it ensures security of data through biometric authentication. It has been developed after performing a detailed case study on healthcare sector in a developing country. Training of the signature samples for authentication purpose has been performed in parallel on hadoop MapReduce framework using Resilient Backpropagation neural network. From rigorous experiments it can be concluded that it achieves a speedup of 9x, Equal error rate (EER) of 0.12, sensitivity of 0.98 and specificity of 0.95 as compared to other approaches existing in literature.

## I. INTRODUCTION

MEDICAL data theft is rising at an alarming rate. Health data is the most vulnerable, easy and tempting target for hackers. According to a survey conducted by Ponemon institute, 2.32 million Americans have been victims to medical data theft [1]. Damages occurring from medical data theft are graver than other type of thefts, since in other types of thefts like credit card or bank ATMs, the bank can be reported of the theft and immediate actions are taken but in health data theft fraud detection is rather tricky and slow. Heath data has become a very tempting target for frauds due to a number of reasons. One of the reasons being the easy access to health data as compared to any other kind of financial data due to lack of security provided in storing the data. Further this information is used by the thieves for creating fraud credit card accounts and file fake tax returns. This health information can be misused to purchase drugs or medical equipments illegally or even to claim false medical care. From past two years, the highest number of hackings is suffered by health industry. According to the Identity Theft Resource Center, it accounted for 43% of the total data breaches recorded [2]. And most of the time the victims are not even aware of the theft. Patient's privacy is first and foremost important issue which must be addressed while switching to e-healthcare system. Thus, a mechanism is required for proper access and retrieval of heath data in a secure manner.

The number of healthcare records is increasing at an exponential rate and it needs to be maintained in an efficient manner. Each of the medical record comprises of many pages comprising of text, images and graphs. Thus, each of these records is in unstructured and unencrypted form. The size of these records is therefore becoming so large that the traditional systems fail to handle and store it. Furthermore, processing of these huge records is an issue. Also, reputed healthcare centers have several branches and there is a need for global accessibility of their medical records. Hence, Cloud computing comes up as a viable solution in this scenario with added advantage of providing these facilities in a scalable, flexible and robust manner, along with location independent access.

Modern technologies are being integrated to already existing medical services. For example EMR (electronic medical records) and EHR (electronic health records) are those two technologies that have been adopted by some of the medical healthcare centers to maintain health records. But

- Kashish Ara Shakil is with the Department of Computer Science, Jamia Millia Islamia, New Delhi 110025. E-mail: shakilkashish@yahoo.co.in.
- Farhana Javed Zareen is with the Department of Computer Science, Jamia Millia Islamia, New Delhi 110025. E-mail: farhanazareen@yahoo.com.
- Mansaf Alam is with Department of Computer Science, Jamia Millia Islamia, New Delhi 110025. E-mail: malam2@jmi.ac.in
- Suraiya Jabin is with Department of Computer Science, Jamia Millia Islamia, New Delhi 110025. E-mail: sjabin@jmi.ac.in
*Equal Contributions

these technologies are susceptible to security threats by hackers and employees of the center. Biometrics based data authentication provides an apt solution for providing security of health data stored in cloud. Biometric security can be obtained from different methods such as confidentiality, integrity, non-repudiation and authentication as shown in fig. 1. Amongst which, authentication is a prime concern so that any illegal access can be prohibited, thus we have focused on biometric authentication in this paper. There are many different biometric authentication mechanisms in practice but biometric signatures are becoming more and more popular because of its social acceptability. Biometric signatures are different from conventional static signatures as static signature verification is based on just the structural properties but biometric signature authentication takes into account the dynamic features of a signature such as velocity, acceleration, total time, pen tilt angles, pen-ups and pen-downs which assures maximum security to the user.

Therefore, an amalgamation of biometrics and cloud can be thought of as an elixir for management of e-health records. In order to speedup processing of data and ensure a secure, scalable storage and management of data cloud computing has been used in combination with biometric authentication mechanism for storage and processing purpose in BAMHealthCloud. It makes use of parallel computing frameworks such as apache hadoop[3] for carrying out the authentication of data and artificial neural network for training and processing of dynamic signature data. Furthermore, priority has also been taken into consideration for defining various levels of data access in order to safeguard high profile clients against data theft which can set grounds for crimes such as defamation of high profile patients through rumors, baseless allegations and falsehood [4].

Thus, this paper proposes BAMHealthCloud and discusses various existing approaches for e- health data and presents a case study about management of health data in a developing country. It suggests BAMHealthCloud as a solution for secure data access, retrieval and management. It has two modules namely health data management system and biometric authentication agent. Following are the contributions of this paper:

- Proposal of BAMHealthCloud which is a novel framework for biometric authentication of data in a healthcare cloud.
- Development of a priority based parallel algorithm ALGOHealthSecurityCheck for providing a secure access to data. Results showed that a speedup of 9x was achieved with the proposed distributed approach as compared to the other approaches existing in literature.
- Validation of BAMHealthCloud through experiments which showed a better performance in terms of EER of 0.12, sensitivity of 0.98 and specificity of 0.95 which is less than the values existing in literature.

The rest of the paper is organized as follows: section II gives the problem description followed by a survey of recent literatures pertaining to health care data in section III. Section IV discusses the case study and proposes BAMHealthCloud framework followed by section V where authentication of the user's access to data is elucidated. Section VI validates the proposed framework through rigorous experimentations which are explained further by results. Finally the paper concludes with the conclusion in section VII.

## II. PROBLEM DESCRIPTION

### A. HealthCare Cloud

Let $H_C$ denote the health care cloud Hc={S, P, D, $S_m$} where S denotes staff i.e. S = {$S_1$, $S_2$, $S_3$, …} where $S_i$ is the $i^{th}$ staff member having access to health data stored in cloud, Pt represents patients i.e. Pt = {$Pt_1$, $Pt_2$, $Pt_3$, …} where $Pt_i$ is the ith patient having access to only his or her record, D denotes data store and $S_m$ security manager. The problem lies in how S and Pt interact with the entity D and how access to D is restricted by the component $S_m$.

### B. Data Management Problem

Health data usually contains records spanning across hundreds of GB's of data. This data is usually difficult to manage using traditional tools and techniques available at disposal. Thus, we need a system that can handle such data characterized by large volume and variety. In order to meet this end the proposed technique uses health data management system which has been built over cloud database management system architecture [5].

### C. Biometric Security management problem

The data stored at a third party location is prone to intrusive attacks, thus we need a management system to secure the system from such attacks. The two main objectives are to provide proper access to legitimate users and to authenticate the users (patients and staffs). The authentication is provided through the biometric authentication agent incorporated in the system. This agent also provides access control to users, so that no outsider or attacker is able to access the system with malicious intentions. All the other symbols used and their explanations is given in table 1.

## III. LITERATURE SURVEY

The healthcare data is growing at an exponential rate, the sources being patient's individual records, data from clinical trials, radiology images and genomics sequences data. It is estimated that this data will escalate to a value of 25,000 petabyte by 2020 [6]. Virtualization and cloud computing are technologies that are gaining popularity for capturing, manipulation and storage of data of large volume. Some of the data management and processing platforms are Hadoop framework and Amazon Corporation EC2. Storing health information in cloud involves its access by a remote desktop, PDAs and mobile devices, but the access through such devices lead to a security issue. The work done in this paper focuses on health care data management through cloud and hadoop distributed programming framework. Furthermore, biometric authentication is used for securing access to this data.

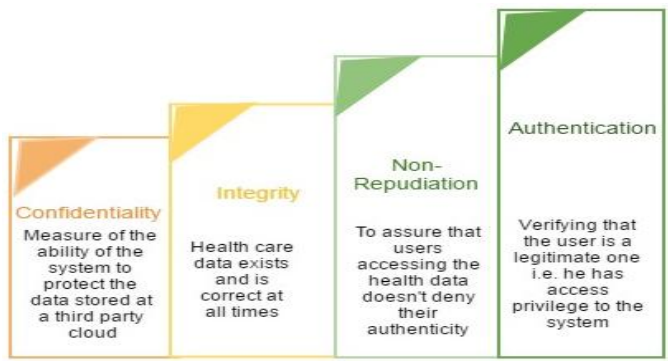

Fig. 1: Different Levels of Biometric Security

According to NIST [7] Cloud computing is defined as a model of ubiquitous computing which provides an on demand access to a set of resources such as network, services and storage in a flexible manner. Some of its benefits include resource pooling, rapid elasticity and on demand services. These are the features that have facilitated the use of cloud computing in various domains including healthcare. Thrust areas of healthcare where cloud computing is becoming popular include DNA sequencing using hadoop for parallel execution of large DNA sequences without any compromise on the accuracy [8]. Biomedical information sharing is another area where cloud computing is gaining popularity for investigators to share data and information with each other but involves risks like data loss and unauthorized access to data [9]. This work makes uses of hadoop for parallel execution in order to achieve a high speed up and robust access to resources along with overcoming the security issue by employing an authentication element to it.

Biometric system can be used for providing authentication in a scenario such as healthcare data storage and retrieval. The main requirement of implementing biometric in healthcare sector is the need for privacy and confidentiality of patient's records. There are some international regulations like HIPAA (Health Insurance Portability and Accountability Act) [10], the Australian Privacy Principles Act and the European Data Protection Directive that demand high level of security, access control, and exchange of sensitive data. Biometric based technology provides a secured method to perform access control and individual identification as compared to other traditional technologies. It is very difficult to reproduce biometric features and therefore it can be used to easily identify an individual.

In [11], authors have used bimodal approach for providing security for electronic medical record, this bimodal approach combines voice recognition and signature authentication technologies. Multi-biometric system integrated with cloud can also provide security for healthcare data [12]. The data in this case is stored in UBUNTU Enterprise Cloud Eucalyptus Database (UEC). In [13] a semi continuous authentication mechanism has been designed for authentication of a layered health monitor framework. It is aimed at reducing misuse of data stored in a heterogeneous cloud system. It uses a fusion of behavioral and physiological biometric i.e. keystroke and face recognition. In BAMHealthCloud, the behavioral as well as the structural characteristics of a person's signature is taken for performing authentication. Furthermore, signatures are highly socially acceptable means of individual authentication. A cloud based approach is used to implement signature based authentication system in order to decrease the running time, and hence BAMHealthCloud achieves high accuracy along with high speedup.

Hence, we have used signature biometric for security of health data stored in cloud. This is the novelty of our paper that we have used behavioral biometric technique like signature which has not been used in the recent literature. It provides minimum error rate and is socially acceptable. Integrating biometric with cloud provides high levels of security which is fast, globally accessible, scalable and robust. In addition, the health care staffs are using hand held devices such as mobile phones, tablets, PDAs to access the records of patients. Using signature biometric amalgamated with cloud technology and integrated with handheld devices would be a foolproof security system for data access and retrieval without any extra hardware cost.

TABLE I
SYMBOLS AND NOTATIONS

| Symbols | Explanations |
|---|---|
| $I_{d^{jk}}$ | Input signature sample, representing $k^{th}$ sample of $j^{th}$ user |
| $O_d$ | Output data after performing MapReduce on Input data |
| $C_{ov}$ | The covariance matrix obtained after performing Sigcovariance function |
| s | Square root of the diagonal elements of the covariance matrix |
| $C_{or}$ | Correlation matrix obtained from covariance matrix and product of square root matrix |
| $P_{cacoef}$ | The matrix representing the PCA coefficients |
| $T_{loc}$ | The local trained network |
| Φ | NULL |
| ∪ | Union |
| $T_{Net}$ | Consolidated trained network |
| $E_{rr}$ | Error of the network |
| $L_{oc}$ | Number of local networks |

IV. CASE STUDY: HEALTHCARE CENTER AT CAPITAL CITY OF A DEVELOPING COUNTRY

In a developing country like India with the population mark reaching second highest in the world. Health is an issue of prime concern as the individuals are the nations building forces. The capital city itself has a population of approximately 1780 lakh [14]. Lakhs of patients visit healthcare centers and corresponding to each patient a record is maintained. According to department of health and family welfare, the department has to cater to the needs of approximately 160 lakh [15] people plus migratory and floating population from neighboring states. With the evolution of digital era a digital copy is kept and since the numbers of patients escalate on a daily basis therefore this has lead to huge volumes of healthcare records. Therefore, it's the need of the hour to handle such voluminous records.

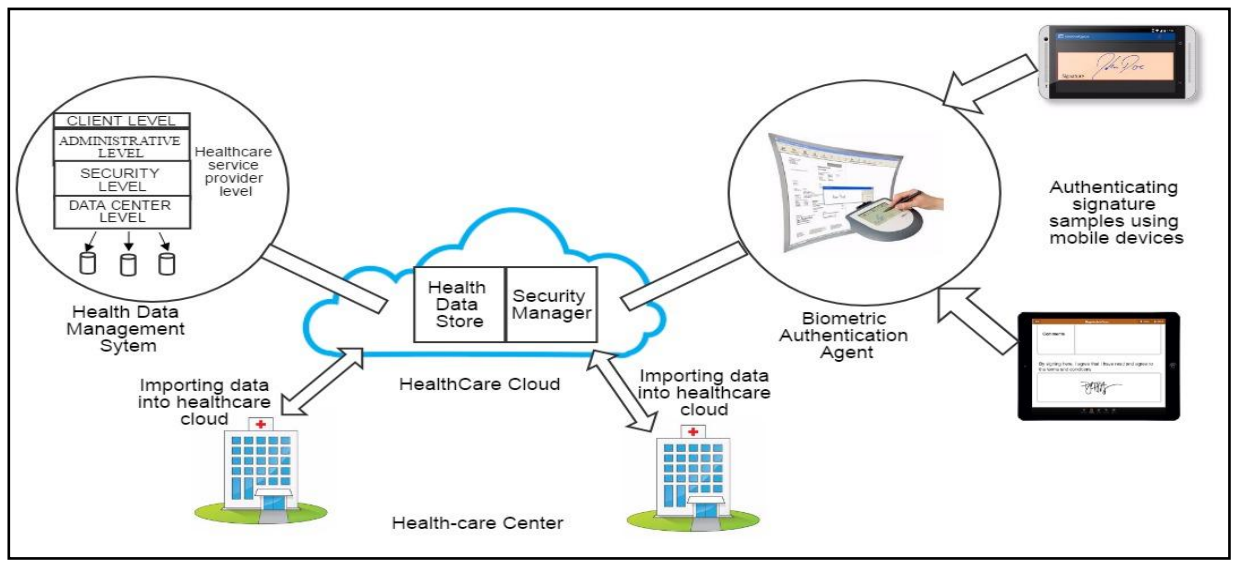

Fig.2: BAMHealthCloud- Cloud based secure biometric architecture

There have been many cases reported in past where healthcare data theft has been performed by employees such as a husband and wife team stole data of 80 patients from Manhattans Lenox Hill Hospital worth more than 14,00000 USD. In another incident a former employee stole fact sheets of 450 patients which can be used for committing medical fraud, tax fraud and health insurance fraud [16].

In this case study, we propose a cloud based solution for handling such huge number of records, the data arising from different health care sources is offloaded to a third party cloud i.e. the proposed healthcare cloud as shown by Fig. 2. The storage and retrieval of this data is handled by the healthcare cloud. The healthcare cloud consists of two components health data store and security manager. The health data store is based on health data management system and security is handled by the biometric authentication agent on behalf of the security manager. The relationship between the different modules is given by Fig. 3. Health data store and security manager are overlapping entities which belongs to the Health Care cloud. The data that is to be secured is stored in Health Data Store and the security manager uses biometric authentication agent for providing the security on cloud.

*A. Health Data Management System*

This component of the healthcare cloud is responsible for management of patients report and other information. It is composed of three layers namely client level, health service provider level and data center level. The health service provider level is further divided into administrative and security level. Biometric authentication agent acts at the security level. Client level provides an interaction interface between the cloud data and its users. The security and resource provisioning is in turn handled by the two components of health service provider level respectively. The data storage and its management is the responsibility of the data center level. Thus, the Health Data Management System is responsible for overall management, storage and retrieval of the healthcare data.

*B. Biometric authentication agent*

This module uses biometric signatures for the purpose of authentication. The dynamic features of a signature is captured using a digitizing tablet which records features like x, y coordinates, velocity of the pen, total time taken to sign, angles of pen while signing, number of pen ups and pen-downs, acceleration etc. After the signature is captured its important features are extracted and then it is preprocessed and stored as a template. After this phase, this template is used for training the data and then these trained networks are stored in cloud. During verification phase, the user's signature is checked against the stored database and it is found out whether the user is genuine or forged.

Moreover, another added advantage of this approach is that the processing is also done on cloud. This saves on the storage and cost along with providing lesser carbon footprints i.e. it's an energy efficient approach as it makes use of mobile and handheld devices like tablets and phones for access purpose.

## V. AUTHENTICATING USER ACCESS

In order to ensure that only legitimate users have access to

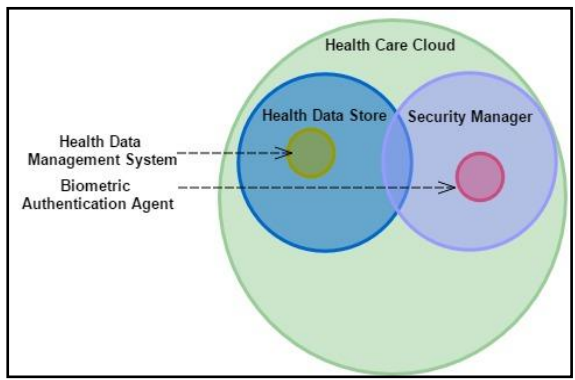

Fig. 3. Relationship between modules of BAMHealthCloud

TABLE II
HEALTH DATA USERS AND THEIR PRIORITIES

| USER TYPES | PRIORITY |
|---|---|
| VIP Patients | 4 |
| Privileged Patients | 3 |
| Privileged Staff | 2 |
| Regular Staff | 1 |

**ALGORITHM 1**
**ALGOHealthSecurityCheck**

**Input:** P: user priority which can be either 1, 2, 3 or 4
1. **Begin:**
2.    **If P equals 1** /* If a user is a regular staff then authentication is performed with low priority*/
3.    **Then**
4.      **ALGOHealthAuthentication**(Priority_Low)
5.    **End If**
6.    **If P equals 2** /* If a user is a privileged staff then authentication is performed with average priority*/
7.    **Then**
8.      **ALGOHealthAuthentication**(Priority_Avg)
9.    **End If**
10.    **If P equals 3** /* If a user is a privileged patient then authentication is performed with high priority*/
11.    **Then**
12.      **ALGOHealthAuthentication**(Priority_High)
13.    **End If**
14.    **If P equals 4** /* If a user is a VIP patient then authentication is performed with very high priority*/
15.    **Then**
16.      **ALGOHealthAuthentication**(Priority_VHigh)
17.    **End If**
18. **End**

**ALGORITHM 2**
**ALGOHealthAuthentication**

**Input:** L: number of users, M: number of signature samples of each user, P: Threshold value taken as input from ALGOHealthSecurityCheck
**Target:** target matrix
**Output:** $\mathcal{T}_{\mathcal{N}et}$ : trained network
1. **Begin:**
2.   **ParFor** j = 1 to L
3.     **ParFor** k = 1 to M
4.       **ReadSamples**($I_{d^{jk}}$) /*Input signature samples from individuals for authentication purpose*/
5.     **End ParFor**
6.   **End ParFor**
7. $\mathcal{O}_d$ = **Sigmapreduce** ($\mathcal{I}_d$, Sigcovariancemapper, Sigcovariancereducer) /* Running Sigmapreduce on input data samples */
8. $\mathcal{C}_{ov}$ = **Sigcovariance**($\mathcal{O}_d$) /*covariance is calculated on the data output from Sigmapreduce */
9. $S$ = $sqrt$(**diagonal**($\mathcal{C}_{ov}$)) /*square root of the diagonal elements of the covariance matrix obtained from previous step is calculated */
10. $\mathcal{C}_{or}$ = $\mathcal{C}_{ov}/S*S'$ /* correlation from covariance matrix and product of square root matrix obtained from step 10 and its transpose is calculated */
11. $\mathcal{P}_{cacoef}$ ← $svd$($\mathcal{C}_{or}$) /* PCA is performed using singular value decomposition on the correlation matrix obtained from step 10*/
12. **Parallelized training:**
13.   **For** j = 1 to L
14.     **For** k = 1 to M
15.       Input1$_{ij}$ ← $\mathcal{P}_{cacoef_{ij}}$ /* inputting preprocessed signature samples obtained from step 11*/
16.     **End** For
17.   **End** For
18. $\mathcal{T}_{\mathcal{N}et}$ = $\Phi$ /*Empty trained network in the beginning*/
19.   **For** l ∈ $\mathcal{L}_{oc}$ **do** /* repeating steps 20-22 for all the local networks*/
20.     $\mathcal{T}_{loc}$ = **netcreate**() /*Creating local networks using Resilient backpropagation algorithm on a feedforward neural network in a distributed manner */
21.     ($\mathcal{T}_{loc}$, $\mathcal{E}_{rr}$) = **Sigtrain**($\mathcal{T}_{loc}$, input, target, P) /* Performing training on local networks created in step 20*/
22.     $\mathcal{T}_{\mathcal{N}et}$ ← $\mathcal{T}_{loc}$ ∪ $\mathcal{T}_{\mathcal{N}et}$ /* Combining all the local networks to form a combined network TNet */
23.   **End** For
24. **End of parallelized training**
25. **Return** $\mathcal{T}_{\mathcal{N}et}$ /* a combined network TNet is returned at the end of parallelized training*/
26. **End**

the data stored on healthcare cloud we use the proposed biometric authentication technique. Moreover, in order to speedup the processing we use parallelized training through MapReduce programming which is based on Hadoop framework. Hadoop is a popular technique for parallel processing of large scale data on cloud. Furthermore, the use mobile phones for authenticating the access to healthcare data ensure that the entire process is energy as well as cost efficient.

The proposed approach is divided into two phases: Phase I-Enrollment phase and Phase II authentication phase

*A. Phase I-Enrollment phase*

In this phase, as shown in Fig. 4, the staff and the patients enrolled in the health care centre are asked to enroll themselves by giving their signature samples using either the signature capturing device or their smart phones that are installed with the signature capturing software. Once a user's signature is captured, the quality of the given signature is checked using SigQuality checker software. The function of this software is to ensure that the recorded signature samples match up to the quality standards required for authentication. Once this match is performed, the features of the samples are extracted and are stored in the healthcare cloud.

*B. Phase II authentication phase*

In this phase as described in Fig. 5, user is checked for his authenticity. The user is asked to give his signature, which is sent for quality check and after that, its features are extracted. Then, this signature sample is checked against the stored user template in the healthcare cloud. This matching is performed using resilient backpropagation algorithm in a feedforward neural network (explained later). The matching process is done in order to find out whether the user is a genuine one or not. This user authentication is done in a parallelized way. ALGOHealthSecurityCheck describes the approach adopted for performing security check on health data lying in the healthcare cloud. In order to check the level of access provided to a user an access hierarchy is defined. This hierarchy is based on the priority of the users. This priority value is assigned at the time of enrollment on a scale of 1-4, 4 is the highest priority value and 1 is the lowest value. The different types of users and their respective priority values are illustrated further in Table. II. Higher the priority value associated with the user more is the level of security guaranteed by the system. This is taken care of by associating a threshold value in the training algorithm i.e. algorithm 1 that depends upon the priority P. According to this algorithm authentication is performed based on the associated priority value. If a user with low priority i.e. a regular staff member accesses the data then its authentication performed with its threshold value set to Priority_Low (lines 1-4). Similarly, the threshold values are set as Priority_Avg, Priority_High and Priority_VHigh for privileged staff, privileged patients and VIP patients respectively.

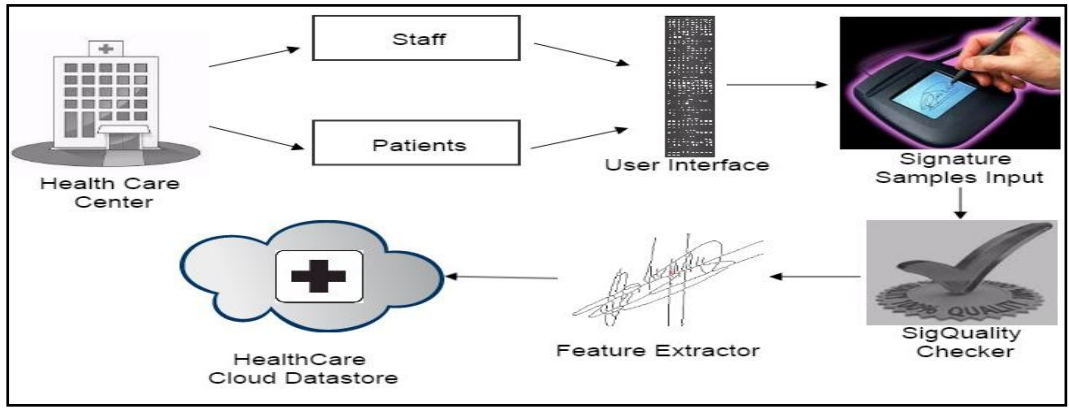

Fig. 4. Enrollment Phase

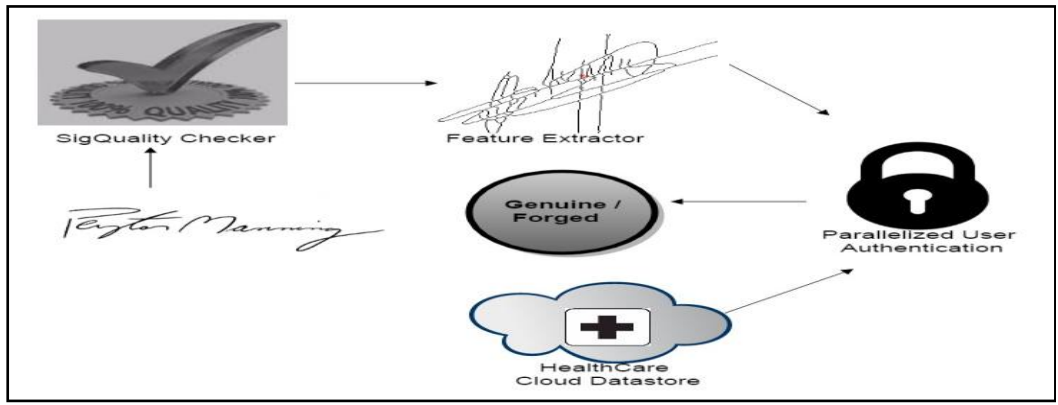

Fig. 5. Authentication Phase

Authentication of the health data based on the priority as threshold value is done by algorithm 2 i.e. ALGOHealthAuthentication. In steps 2 to 6, the signature samples are read. This read is performed in a parallel fashion. $I_{d^{jk}}$ is the input signature sample matrix which represents the $k^{th}$ samples of $j^{th}$ user. After this mapreduce function is applied for calculating the covariance on the input data sample. The usage of mapreduce framework ensures that all the processing of data is done in a parallelized approach on commodity hardware across a distributed cluster. This speeds up the processing and is cost effective along with being fault tolerant.

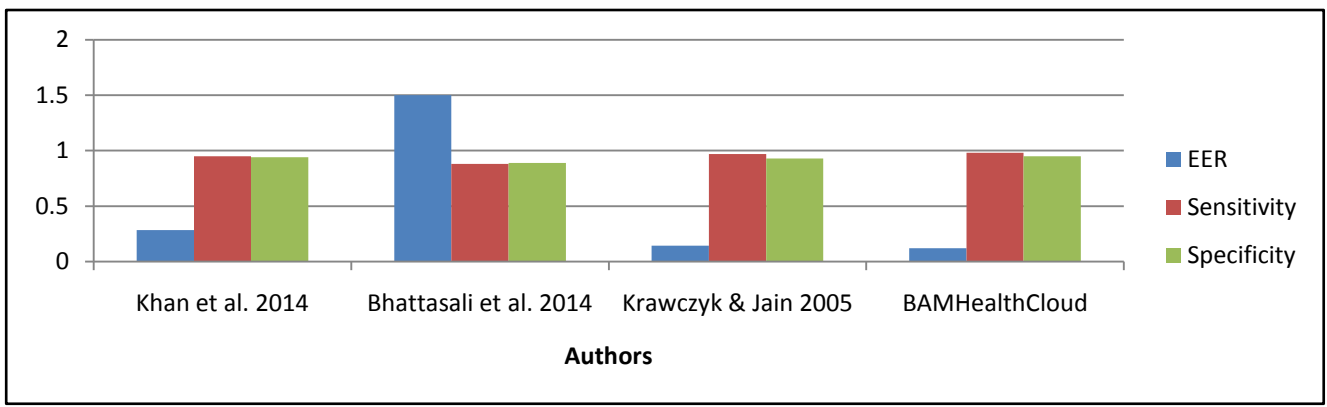

Fig. 6: EER, Sensitivity and specificity of BAMHealthCloud versus approaches used in literature

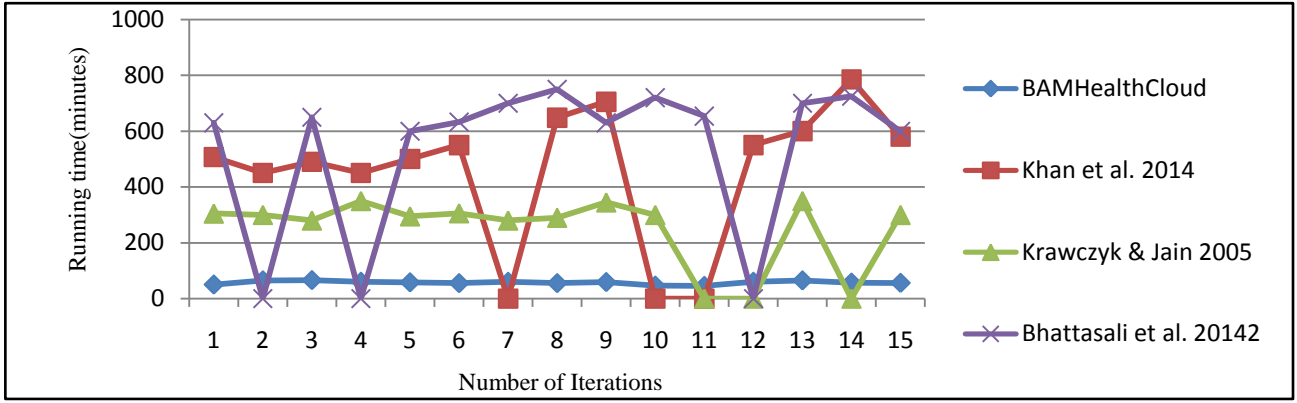

Fig. 7: Running time of BAMHealthCloud versus approaches used in literature

In step 11, the PCA feature extraction is performed on the input data. After this, training of dataset is done using resilient backpropagation algorithm in a parallelized manner.

Resilient backpropogation (RPROP) is a popular algorithm for backprpogation training of multilayer feed-forward neural networks. Backpropagation algorithm involves application of chain rule repeatedly to findout the impact of each weight in accordance to error function E [17].

$$\frac{\partial E(t)}{\partial \omega_{ij}} = \frac{\partial E}{\partial s_i} \cdot \frac{\partial s_i}{\partial net_i} \cdot \frac{\partial net_i}{\partial w_{ij}} \quad (1)$$

where, $\omega_{ij}$ is the weight from neuron j to neuron i. $s_i$ is the output and $net_i$ is the sum of inputs of neurons i.

$$w_{ij}(t+1) = w_{ij}(t) - \epsilon \frac{\partial E}{\partial w_{ij}}(t) \quad (2)$$

RPROP solves the problem of "vanishing gradients", according to which with increase in depth and complexity of an artificial neural network, the gradient propagates backwards by stochastic gradient descent (SGD) backpropagation becomes increasingly smaller which leads to negligible updates in weight along with slow training [18] [17]. It achieves this by using a fixed update value $\delta_{ij}$ whose value can be increased or decreased with each iteration $\eta_+$ and $\eta_-$ multiplicativley. where, $\eta$ is an asymmetric factor. The increase or decrease in values depends on the change in sign of gardient with respect to weight $\omega_{ij}$ It converges to a local minima but is able to skip through flat surfaces fast due to $\eta+$ multiplicative factor. Algorithm 3 shows the pseudocode of Resilient backpropogation algorithm [18]. Here, $\Delta_{ij}$ is the update value for each weight i.e. the size of weight update. $\Delta_{max}$ and $\Delta_{min}$ are the maximum and minimum change and $\Delta w_{ij}$ is the change in weights.

Local networks ($\mathcal{N}_{oc}$) are created using netcreate () function and then these local networks are combined to form a consolidated network of local networks ($\mathcal{TN}_{et}$).This consolidated network is the output of the algorithm.

*Definition 1: netcreate function implements Resilient backpropagation algorithm on a feed forward network in a distributed manner and this distribution is done based on the size of the dataset being used.*

*Definition 2: Sigtrain is the function which takes input and target matrices and trains a network based on the architecture of a local network which was created using Resilient backpropagation algorithm.*

**Theorem 1:** *The time complexity of ALGOHealthSecurityCheck is O(L xM).*

**Proof:** The time complexity for performing security check on health data stored on cloud is O (L x M). It takes O (L x M) for performing security check on low priority users (line 4, ALGORITHM 1) as given in theorem 2. It takes O (L x M) for performing security check on average priority users (line 8, ALGORITHM 1). It takes O (L x M) for performing security check on high priority users (line 12, ALGORITHM 1) and takes O (L x M) for performing security check on very high priority users (line 16, ALGORITHM 1).Thus, the complexity of ALGOHealthSecurityCheck is maximum (O (L x M), O (L x M), O (L x M), O (L X M)) i.e. O (L x M)

**Theorem 2:** *The time complexity of ALGOHealthAuthentication is O (L x M) where L is the number of users and M is the number of signature samples corresponding to each user.*

**Proof:** For performing authentication of health data the complexity is *O (L x M)*. It takes *O (L x M)* for performing read on sample data *(line 4,* ALGORITHM 2), It takes O (L *x M)* for performing mapreduce function on data samples *(line 7,* ALGORITHM 2*)*. Sigcovariance finds the covariance in O ($L^2$) time and steps 13-17 take O (LxM) time and steps 20-21 takes O(LxM) time. Thus, the complexity of ALGOHealthAuthentication is maximum (O (L x M), O (L x M), O ($L^2$), O (L x M), O (L x M))

TABLE III
FEATURES RECORDED WHILE SIGNING [19]

| Parameters | Features | Unit |
|---|---|---|
| Acceleration | $X_\alpha$ | $m/s^2$ |
| | $Y_\alpha$ | $m/s^2$ |
| | $Z_\alpha$ | $m/s^2$ |
| Magnetic Field | $X_\mu$ | Mt |
| | $Y_\mu$ | Mt |
| | $Z_\mu$ | Mt |
| Orientation | Azimuth | Degrees |
| | Pitch | Degrees |
| | Roll | Degrees |
| Angular Velocity | $X_v$ | rad/s |
| | $Y_v$ | rad/s |
| | $Z_v$ | rad/s |

## VI. EXPERIMENTS AND RESULTS

### A. Experimental Setup

#### 1) System setup

An 8 cabinet cluster with 397 nodes was taken for carrying out experimentations of BAMHealthCloud. Each node had 12 cores and hence, the total number of cores was 4764 (397x 12). The total memory available per node was 64 GB and therefore, the total memory available was 768GB (12x 64). The experimental data that were used are real handwritten signature data that were collected through handheld devices.

#### 2) Dataset Used

Different handheld devices such as Smart phones, tablets, phablets, personal digital assistants (PDAs) were used for acquiring data. The signature samples were collected from 9000 users. The users involved both the staff and patients. The data has been recorded in five temporal sessions separated over a period of one week per 1500 users. The data collected comprised of 2000 staff members and 7000 patients. 40 samples were collected from each user, 25 were genuine and 15 samples were forged. Therefore, the total number of samples stored was 360000.

In order to comply to the ethics involved with collection of such significant and critical information, the participants consent was taken on the clause that their identity will be kept anonymous. The participants were only identified as either staff or patients, apart from this all the personal information were kept anonymous.

Table III shows the features that have been recorded while capturing signatures [19]. The features include Acceleration, Magnetic Field, Orientation and Angular Velocity which were recorded while capturing signatures.

#### 3) Data Preparation and Processing

In order to prepare data for training purpose, the raw data was firstly preprocessed by using hadoop framework. The data was first imported into hadoop distribuited file system and was then reduced using Principal component analysis (PCA) [20] technique. MapReduce code for PCA was executed in order to carry out all the computations in a parallel manner. It should be noted that since the dataset used was very large, the processing involved multiple mapper functions spanning across several nodes of the cluster. Hadoop2.x which is the latest version has been utilized in our experimentations. The replication factor of 3 which was adopted ensured that our system is robust against any node failures. Furthermore, robustness was also ensured through speculative execution property of hadoop cluster.

### B. Evaluation metrics

Following metrics have been used to evaluate the framework

1) *Sensitivity:*
It is the probability of a system to correctly classify that a signature sample belongs to a particular authorized health data user and is given by equation 1.

$$Sensitivity = \frac{TrueGenuine}{TrueGenuine + FalseForged} \quad (1)$$

2) *Specificity:*
It is the probability of a system to correctly classify that the particular sample does not belong to the user class and is given by equation (2).

$$Specificity = \frac{TrueForged}{TrueForged + TrueForged} \quad (2)$$

3) *Equal Error Rate (EER):*
i) *False acceptance rate (FAR)* measures the rate of the system that how many times a system accepts a forged sample as correct input. FAR can be represented using equation 3.

$$FAR = \frac{FalseGenuine}{FalseGenuine + TrueForged} \quad (3)$$

ii) *False rejection rate (FRR)* measures the rate of the system that how many times a system rejects a genuine sample as forged. FRR can be represented using equation 4.

$$FRR = \frac{FalseForged}{FalseForged + TrueGenuine} \quad (4)$$

iii) *EER* is the value when both FAR and FRR becomes equal and is represented using equation 5.

$$EER = FAR|_{FAR=FRR} = FRR|_{FRR=FAR} \quad (5)$$

---

**ALGORITHM 3**
**RESILIENT BACKPROPAGATION ALGORITHM[18]**

1. $\eta_+ = 1.2$, $\eta_- = 0.5$, $\Delta max = 50$, $\Delta_{min} = 10^{-6}$
2. pick $\Delta_{ij}(0)$
3. $\Delta\omega_{ij}(0) = -\text{sgn}\frac{\partial E(0)}{\partial w_{ij}} \cdot \Delta_{ij}(0)$
4. **for all** $t \in [1..T]$ **do**
5.    **if** $\frac{\partial E(t)}{\partial \omega_{ij}} \cdot \frac{\partial E(t-1)}{\partial \omega_{ij}} > 0$ **then**
6.       $\Delta_{ij}(t) = \min\{\Delta_{ij}(t-1) \cdot \eta_+, \Delta_{max}\}$
7.       $\Delta w_{ij}(t) = -\text{sgn}\frac{\partial E(t)}{\partial w_{ij}} \cdot \Delta ij(t)$
8.       $w_{ij}(t+1) = w_{ij}(t) + \Delta w_{ij}(t)$
9.       $\frac{\partial E(t-1)}{\partial w_{ij}} = \frac{\partial E(t)}{\partial w_{ij}}$
10.   **else if** $\frac{\partial E(t)}{\partial w_{ij}} \cdot \frac{\partial E(t-1)}{\partial w_{ij}} < 0$ **then**
11.       $\Delta_{ij}(t) = \max\{\Delta_{ij}(t-1) \cdot \eta_-, \Delta_{min}\}$
12.       $\frac{\partial E(t-1)}{\partial w_{ij}} = 0$
13.   **else**
14.       $\Delta w_{ij}(t) = -\text{sgn}\frac{\partial E(t)}{\partial w_{ij}} \cdot \Delta ij(t)$
15.       $w_{ij}(t+1) = w_{ij}(t) + \Delta w_{ij}(t)$
16.       $\frac{\partial E(t-1)}{\partial w_{ij}} = \frac{\partial E(t)}{\partial w_{ij}}$
17.   **end if**
18. **end for**

Where, TrueGenuine is the number of genuine users who are classified as a genuine by BAMHealthCloud. FalseGenuine is number of genuine users who have not been correctly classified, TrueForged is number of forged users who have been falsely classified as genuine and FalseForged is number of forged users who have been correctly classified as forged.

4) *Speedup:*

Speedup of BAMHealthCloud is defined as the amount of performance gain which is obtained by running the system on a single machine in contrast to that when it is run on multiple machines. The speedup according to Amdahl's law [21] is given by equation 6.

$$S(P) = \left(\frac{T(1)}{T(P)}\right) \quad (6)$$

Where P is the number of processors, S (P) is the speedup achieved while running computation on P processors, T (1) is the running time on a single processor and T (P) is the running time on P processors.

*C. Results*

In order to validate the authenticity of BAMHealthCloud, experiments have been performed. The results of BAMHealthCloud have been compared to the other approaches that have been used in recent literature and are summarized in Fig. 6 and Fig. 7. Fig. 6 shows the results in terms of Equal Error Rate (EER), sensitivity and specificity, while Fig. 7 shows the plot of running times of the different approaches.

From Fig. 6, we can see that results obtained from the proposed framework are better than the other compared approaches in literature. BAMHealthCloud achieves an EER of 0.12 which is better than the rest of the approaches used. The sensitivity of the proposed system is 0.98 and the specificity is 0.95 which is again better than the existing approaches implemented in literature.

In order to further validate the effectiveness of the proposed approach, the running time of BAMHealthCloud has been calculated along with running time of other sequential approaches proposed in [12], [13] and [11] as shown in Fig.7. The experiments have been conducted rigorously 15 times and from the results it can be concluded that BAMHealthCloud took lesser time to run than these approaches. The results furthermore, show that a speedup of 9x was achieved by BAMHealthCloud using equation 6.

With the achieved results showing EER of 0.12, sensitivity of 0.98, specificity of 0.95 and speed up of 9x, it can be inferred that BAMHealthCloud outperforms all the other methods that have been used in healthcare sectors. Moreover, this approach can be successfully deployed for securing health data in real life.

VII. CONCLUSION

In order to handle ever growing data of health sectors and to provide security to different users and staff involved, a cloud based biometric authentication system BAMHealthCloud has been proposed in this paper. This model is developed after performing a detailed case study on a healthcare centre in a developing country. This model has two different components, one component takes care of the management of the huge data that is being generated on a daily basis and other one takes care of the security aspect. ALGOHealthSecurityCheck has been proposed in this work which performs security check using parallelized MapReduce programming model. The use of this model ensures the scalability, flexibility and robustness of the system. Experiments performed on this system reveal that BAMHealthCloud achieves a speedup of 9x which is better than other systems implemented in recent works. Priority based system has been incorporated in BAMHealthCloud which results in fast fetching and processing of crucial records. The system achieves an EER of 0.12, sensitivity of 0.98 and specificity of 0.95 which is comparable to other approaches existing in literature.

BAMHealthCloud can be successfully deployed in other sectors as well such as defense, banking sector, organizations, and educational sectors. It is well suited for applications of any size as it offers horizontal scaling of resources and thus is suitable for big data processing.

ACKNOWLEDGMENT

This is to acknowledge that Farhana Javed Zareen is the corresponding author of this paper. It is further acknowledged that Farhana Javed Zareen and Kashish Ara Shakil share equal contributions for the work carried out in this article. This work was supported by a grant from "Young Faculty Research Fellowship" under Visvesvaraya PhD Scheme for Electronics and IT, Department of Electronics & Information Technology (DeitY), Ministry of Communications & IT, Government of India.

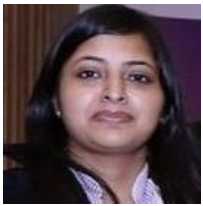
Kashish Ara Shakil has received her Bachelor's degree in Computer Science from Delhi University and has an MCA degree as well. She is currently pursuing her doctoral studies in Computer Science from Jamia Millia Islamia (A Central University). She has written several research papers in the field of Cloud computing. Her area of interest includes database management using Cloud computing, distributed and service computing.

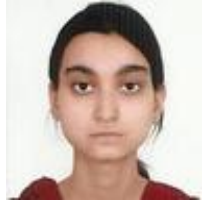
Farhana Javed Zareen, Ph.D. Scholar, Jamia Millia Islamia, Central University, New Delhi, India. She is working in the field of biometric authentication from past 3 years. She has received her bachelor's degree (2010) and master's degree (2012) in computer science from Calcutta University. She has published two research papers in pattern recognition field. Her research interests include pattern recognition, biometric authentication, image processing and machine learning.

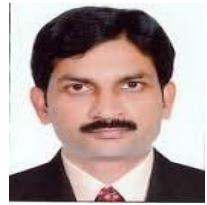
Mansaf Alam received his doctoral degree in computer Science from Jamia Millia Islamia, New Delhi in the year 2009. He is currently working as an Assistant Professor at the Department of Computer Science, Jamia Millia Islamia. He is also the Editor-in-Chief, Journal of Applied Information Science. His areas of research include Cloud database management system (CDBMS), Object Oriented Database System (OODBMS), Genetic Programming, Bioinformatics, Image Processing, Information Retrieval and Data Mining.

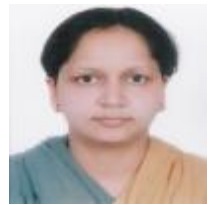
Suraiya Jabin is an Assistant professor in the Department of Computer Science, Jamia Millia Islamia, Central University, New Delhi, India. She received her Ph. D degree in 2009 from Department of Computer Science, Hamdard University, and New Delhi, India. Her research interests include Artificial Intelligence, Pattern Recognition, Soft Computing, and Biometrics.